\documentclass[prl,reprint,showpacs]{revtex4-1}
\usepackage{amsfonts}
\usepackage{amssymb}
\usepackage{amsmath}
\usepackage{graphicx}

\begin{document}
\title{Eigenfunction distribution for the Rosenzweig--Porter model}
\author{E. Bogomolny}
\affiliation{CNRS, Universit\'e Paris-Sud, UMR 8626,
Laboratoire de Physique Th\'eorique et Mod\`eles Statistiques, 91405 Orsay,
France}
\author{M. Sieber}
\affiliation{School of  Mathematics, University of Bristol, University Walk, Bristol BS8 1TW, UK}
\begin{abstract}
The statistical distribution of eigenfunctions for the Rosenzweig-Porter model is derived for the region where eigenfunctions have fractal behaviour. The result is based on simple physical ideas and leads to transparent explicit formulas which agree very well with numerical calculations. It constitutes a rare case where a non-trivial eigenfunction distribution is obtained in a closed form.  
\end{abstract}
\maketitle

Random matrix theory has been successfully applied to a vast number of different problems ranging from nuclear physics to number theory (see eg. \cite{mehta,rmthandbook} and references therein). Recently \cite{kravtsov} it was demonstrated that it is also applicable for describing models with fractal eigenfunctions. The difference of the model considered in \cite{kravtsov} from other models with fractal eigenfunctions, such as the power-law random banded matrices \cite{mirlin} and the ultrametric matrices \cite{ossipov}, 
is that in the latter models fractality (or even multifractality) exits  only  at special critical values of the parameters, but in the former model fractality has been observed in  part of  the whole delocalised phase.  

The model in question belongs to  the Rosenzweig-Porter  (RP) matrix ensembles $H_{ij}$  \cite{rp} where all matrix elements are independent (up to the Hermitian symmetry) Gaussian variables with zero mean, and whose variances of diagonal and off-diagonal elements depend on different powers of the matrix dimension $N$
 \begin{equation}
 \langle H_{i j}\rangle=0, \quad \langle H_{i i}^2\rangle=1,\quad 
\langle H_{i j}^2\rangle_{i\neq j}= \frac{\epsilon^2}{N^{\gamma}}  
\label{matrix}
 \end{equation}
where $ i, j=1,\ldots, N$ and $\epsilon$ is a constant. 

For clarity we consider real symmetric matrices (of  GOE-type). Generalisation to other symmetry classes is straightforward. 

It has been established (with physical rigour in \cite{pandey}-\cite{kunz} and proved mathematically in \cite{soosten}) that when $\gamma>2$ all states in the model are localised and the spectral statistics is Poissonian. When $\gamma<1$ after rescaling one gets the usual random matrix ensembles, therefore all states are delocalised and the spectral statistics coincides with GOE (mathematically it follows from the results of \cite{yau}).  

In \cite{kravtsov} the remaining interval $1<\gamma<2$ has been thoroughly investigated and it was demonstrated that the eigenfunctions are delocalised but have unusual fractal properties. In particular, eigenfunction moments
\begin{equation}
I_q=\langle \sum_j |\Psi_j|^{2q} \rangle
\label{definition}
\end{equation}
for $q>\frac{1}{2}$ scale with a non-trivial power of $N$
\begin{equation}
I_q\underset{N\to \infty}{\longrightarrow} N^{-(q-1)D_q} \, C_q,\qquad D_q=2-\gamma\ . 
\end{equation}
Recently, the existence of fractal states in this model has been rigorously proved \cite{warzel}.

The purpose of this note is to obtain an exact distribution of eigenfunctions,  $\Psi_i(E_\alpha)$, in the RP model \eqref{matrix} 
\begin{equation}
\sum_{j=1}^N H_{ij}\Psi_j(E_\alpha)=E_{\alpha}\Psi_i(E_\alpha) .
\end{equation}
for  large but finite matrix dimensions.

Our derivation is based on two (heuristic) statements. The first is related to the form of the mean value of the modulus  square of eigenfunction components for large $N$
\begin{equation}
\Sigma_j^2(E)\equiv \langle |\Psi_j(E)|^2\rangle \approx \frac{C^2\  \Gamma(E) }{\pi \rho(E)N ((E-e_j)^2+\Gamma^2(E))} .
\label{cauchy}
\end{equation}
 Here  the average is taken over off-diagonal matrix elements taking the diagonal elements $e_j\equiv H_{jj}$ fixed. The width  $\Gamma(E)$  is called the spreading width, and for large $N$ it is given by the Fermi golden rule
 \begin{equation}
 \Gamma(E)= \frac{\pi \epsilon^2}{N^{\gamma-1}} \rho(E) 
 \label{fermi}
 \end{equation}
 where $\rho(E)$ is the normalised level density of the matrices \eqref{matrix}. For large $N$ and $\gamma>1$ it is equal to the density of the diagonal elements
 \begin{equation} \label{dens}
 \rho(E)=\frac{1}{\sqrt{2\pi}}\exp \left (-\frac{E^2}{2} \right ) .
 \end{equation}
 The value of constant $C$ depends on the chosen normalisation of the eigenfunctions. Usually eigenfunctions are normalised as follows
 $\sum_j|\Psi_j(E_\alpha)|^2=1$ or 
 \begin{equation}
\sum_{\alpha}|\Psi_j(E_\alpha)|^2=1 \longrightarrow  \int \rho(E) \langle |\Psi_j(E)|^2\rangle \mathrm{d}E=\frac{1}{N}\ .
 \label{normalisation}
 \end{equation}
 In this case $C=1$.
 But  we shall see  that it is convenient to consider the statistical distribution not of $\Psi_i$ itself but of the variable $y=C\Psi_i$ where the constant $C$ is a certain power of $N$. This is equivalent to choosing a different normalisation of eigenfunctions.    
 
 The (probably) simplest way to get the result \eqref{cauchy} is to use a recursive relation for the Green function $G=(E-H)^{-1}$.  Fixing the diagonal element $e_i=H_{i i}$ and expanding the determinant over column and row $i$ one gets the identity (called in the mathematical literature the Schur complement formula)  with $z=E-\mathrm{i}\eta$ and $\eta \to 0_+$
 \begin{equation}
 G_{ii}(z)=\Big( z-e_i-\sum_{j,k\neq i} H_{ij}G_{jk}^{(i)}(z)H_{ki} \Big)^{-1}
 \end{equation}
 where $G^{(i)}(E)$ is the Green function of the matrix obtained from $H$ by removing the row and column $i$.  
 
 The next approximations seem natural and can be rigorously proved in certain cases. First, one takes into account only diagonal terms in the double sum and substitutes random matrix elements by their expectation values
 \begin{equation}
 \sum_{j,k\neq i} H_{ij}G_{jk}^{(i)}H_{ki}\approx  \frac{\epsilon^2}{N^{\gamma-1}}\tilde{G}^{(i)},\quad \tilde{G}^{(i)}=\frac{1}{N-1}\mathrm{Tr}\, G^{(i)} .
 \end{equation}
 Second, for large  $N$  one can ignore small contributions of off-diagonal elements to ${G}^{(i)}$ and use instead the free diagonal Green function. Using the self-averaging property of this quantity one gets the usual result
 \begin{equation}
 \tilde{G}^{(i)} \underset{N\to\infty}{\longrightarrow}\int \frac{\rho(e)\mathrm{d}e}{E-\mathrm{i}\eta-e}\underset{\eta\to 0_+}{\longrightarrow} 
 \mathrm{pv}\int \frac{\rho(e)\mathrm{d}e}{E-e} +\mathrm{i}\pi \rho(E)
 \label{green}
  \end{equation}
 where $\rho(e)$ is density \eqref{dens} of diagonal entries of the matrices \eqref{matrix} and  pv denotes the principal value of the integral.
 
 The first term in \eqref{green} after multiplication by $N^{1-\gamma}$ gives only a small energy shift in Eq.~\eqref{cauchy}  (when $\gamma>1$) and will be ignored in what follows.  Taking only the imaginary part gives 
 Eq.~\eqref{cauchy}. Notice that in the chosen approximation the normalisation \eqref{normalisation} is fulfilled.
 
 The appearance of the characteristic Breit-Wigner shape \eqref{cauchy} in the case when the interaction between unperturbed levels is small is well known  and was observed in many different settings. It was Wigner \cite{wigner} who proved (for a different model) that in such a case the mean square modulus of eigenfunction components has the form \eqref{cauchy}.  Later this approach was widely used in nuclear physics  and quantum chaos (see e.g. Refs. \cite{bm}--\cite{borgonovi} besides others).  The notion of spreading width $\Gamma$ by itself is very useful in applications. The point is that the ratio between it and the mean level spacing determines between  how many levels the initially localised state spreads after the interaction is switched on. Therefore without further calculations it is physically obvious that for an interaction as in \eqref{fermi} and level spacings of the order of $1/N$ (for $\gamma>1$) an exact eigenfunction is spread between $N^{2-\gamma}$ levels when $1<\gamma<2$ and will be fully localised when $\gamma>2$. Notice that the exact result \cite{bogomolny} for the case of usual random matrix models perturbed by rank-one perturbations leads to similar formulas.
 
 The second important ingredient of our derivation is the assumption that the distribution of eigenfunctions with fixed diagonal elements can be well approximated by a Gaussian function with zero mean and the variance given by  Eq.~\eqref{cauchy}
 \begin{equation}
 P(\Psi_j(E))=\frac{1}{\sqrt{2\pi \Sigma_j^2(E)}}\exp \left ( -\frac{|\Psi_j(E)|^2}{2\Sigma_j^2(E)}\right )\ .
 \label{local_PT}
 \end{equation}
 Such a simple assumption (a local Porter-Thomas law) has been used  for many different problems (see e.g. \cite{zelevinsky}, \cite{izrailev}, \cite{borgonovi}), and it has been seen as a necessary condition to get (with physical rigour) the thermalisation  from quantum mechanics \cite{deutsch}. Recently this property has been rigorously proved for the RP model \cite{benigni}.
 
 The final step to find the statistical distribution of eigenfunctions consists in averaging \eqref{local_PT} over energy $e_j$. It gives ($x=\Psi_j(E)$)
 \begin{equation}
 P(x)=\int   \frac{\rho(e)}{\sqrt{2\pi \Sigma_j^2(E)}}\exp \left ( -\frac{x^2}{2\Sigma_j^2(E)}\right )\mathrm{d}e .
 \end{equation}
 Substituting the above values results in
 \begin{widetext}
 \begin{equation}
 P(x)=\frac{1}{2\pi \sqrt{a}}\int_{-\infty}^{\infty}\sqrt{(E-e)^2+\Gamma^2(E)}\, 
 \exp \left ( -\frac{ x^2}{2a}\Big ((E-e)^2+\Gamma^2(E) \Big ) -\frac{e^2}{2}\right ) \mathrm{d}e 
 \label{distribution}
 \end{equation}
 \end{widetext}
 where we introduced the notation 
 \begin{equation}
 a=\frac{C^2\Gamma(E)}{\pi \rho(E) N}=\frac{C^2\epsilon^2}{N^{\gamma}}\ .
 \end{equation}
 This formula gives the distribution of eigenfunctions with  energies in a small window around $E$. The simplest  case corresponds to the centre of the spectrum, $E=0$. The remaining integral can easily be calculated and for $E=0$ one gets
 \begin{equation}
 P(x)_{E=0}= \frac{\delta^2}{4\pi \sqrt{a}} \big [ K_0(\zeta)+K_1(\zeta)\big ]\mathrm{e}^{-\zeta+\frac{\delta^2}{2}}, 
  \label{main}
  \end{equation}
  where 
  \begin{equation}
  \delta\equiv \Gamma(0)=\frac{\sqrt{\pi}\, \epsilon^2}{\sqrt{2}\, N^{\gamma-1}}, \quad \zeta=\frac{\delta^2}{4a} (x^2+a)
  \end{equation}
   and $K_0(z)$ and $K_1(z)$ are  the K-Bessel functions (see e.g. \cite{be}, 7.12 (21))
   \begin{equation}
   K_{\nu}(z)=\int_0^{\infty}\cosh(\nu t)\mathrm{e}^{-z\cosh t}\mathrm{d}t\ .
   \end{equation}
   Formula \eqref{main} is the main result of this note. It represents the distribution of eigenfunctions for the RP model for large matrix dimensions in the region $\gamma>1$ at a small interval around $E=0$. It is straightforward to get a more general expression valid in 
   a finite energy interval  but the result is  cumbersome without producing any new insights.

  It is clear that the bulk contribution corresponds to values of $x$ of the order of $\sqrt{a}$. To clearly see this region it is convenient to use variable
  \begin{equation}
  y=N^{\gamma/2}\ \Psi_j(E), \quad \langle y^2 \rangle=N^{\gamma-1}, \quad |y|\leq N^{\gamma/2}.
  \label{xn}
  \end{equation}
  This choice corresponds to $C=N^{\gamma/2}$ and $a=\epsilon^2$. Because $\delta$ is always small in the large $N$ limit, one can expand Eq.~\eqref{main} with $x=\mathcal{O}(1)$ for $\delta\to 0$. As $K_1(z)\to 1/z$  when $z\to 0$ one gets 
 \begin{equation}
 P(y)_{\textrm{bulk}}= \frac{\epsilon}{\pi (y^2+\epsilon^2)}\ .
 \label{bulk}
 \end{equation}
  The leading correction to this limit is of the order of $\delta^2\ln \delta^2$ and it comes from the expansion of $K_0(\zeta)$.
  
 To  investigate the behaviour of the eigenfunction distribution for   large $x$ (finite values of $\zeta$ in \eqref{main}) it is useful to rescale eigenfunctions as follows
 \begin{equation}
 z=N^{1-\gamma/2}\, \Psi_j(E),\quad \langle z^2\rangle =\frac{1}{N^{\gamma-1}}, \quad |z|\leq N^{1-\gamma/2}\ .
 \label{zn}
 \end{equation}
 This normalisation corresponds to $C=N^{(2-\gamma)/2}$ and $a=\epsilon^2 N^{2-2\gamma}$. Consequently, 
 \begin{equation}
 P(z)_{\mathrm{tail}}= \frac{2\sqrt{2}\, b^3}{\pi \sqrt{\pi}\, N^{\gamma-1}} \big (K_0(b^2 z^2)+K_1(b^2 z^2)\big )\mathrm{e}^{-b^2z^2}
 \label{tail}
 \end{equation}
 with $b=\sqrt{\pi}\,  \epsilon/(2\, \sqrt{2})$.
 When we apply these expansions to eigenfunctions it is necessary to take into account that they lose their validity near the maximum values indicated in  \eqref{xn} and \eqref{zn}. The derivation of large deviation formulas applicable  close to these limits (inherent from the obvious bound $|\Psi_j|\leq 1$) is beyond the scope of this note.  Furthermore, the large $x$ expansion \eqref{zn} and \eqref{tail} does not exist for $\gamma>2$.
 
 Using Eq.~\eqref{main} (or directly from \eqref{distribution}) it is straightforward to calculate moments of eigenfunctions \eqref{definition} in the centre of the spectrum
 \begin{equation}
  I_q=\frac{2^{q-1/2} a^q N \Gamma(q+1/2)}{\sqrt{\pi}\delta^{2q-1}}\Psi \Big (\frac{1}{2},\frac{3}{2}-q;\frac{\delta^2}{2}\Big )
 \end{equation}
 where $\Psi(\alpha,\beta;z)$ is the Tricomi confluent hypergeometric function (see e.g. \cite{be}, 6.5 (2))
 \begin{equation}
 \Psi(\alpha,\beta;z)=\frac{1}{\Gamma(\alpha)}\int_0^{\infty} \mathrm{e}^{-zt}\, t^{\alpha-1}\, (1+t)^{\beta-\alpha-1}\, \mathrm{d}t \ .
 \end{equation}
 If $\beta$ is not an integer this function is a sum of two hypergeometric functions (\cite{be}, 6.5 (7))
 \begin{eqnarray}
 & &\Psi(\alpha,\beta;z)=\frac{\Gamma(1-\beta)}{\Gamma(\alpha-\beta-1)}   {}_1F_1(\alpha;\beta;z)\nonumber\\
 &+&\frac{\Gamma(\beta-1)}{\Gamma(\alpha)}z^{1-\beta}{}_1F_1(\alpha-\beta+1;2-\beta;z)
 \end{eqnarray}
 Going to the limit $\delta\to 0$ one gets (in agreement with \cite{kravtsov})
 \begin{equation}
 I_{q}=N^{-\tau(q)}\, C_{q},\; \tau(q)=\left \{ \begin{array}{cc}\gamma q-1,&q<\tfrac{1}{2}\\(q-1)(2-\gamma),&q>\tfrac{1}{2}\end{array}\right . 
  \end{equation} 
  where pre-factors $C_q$ have the following values
 \begin{eqnarray}
 C_{q<\tfrac{1}{2}}&=&\frac{\epsilon^{2q}}{\pi}\Gamma(q+1/2)\Gamma(1/2-q)\ , \label{small_q}\\
 C_{q>\tfrac{1}{2}}&=& \frac{\Gamma(q-1/2)\Gamma(q+1/2)}{\pi \, b^{2q-2} \, 2^{q-2}\, \Gamma(q)}\ . \label{large_q}
 \end{eqnarray}
It is clear that these asymptotic values correspond to the moments of  distributions \eqref{bulk} and \eqref{tail} respectively. It is also possible to find corrections to the above results (cf. \cite{be}, 6.8). In particular for $-\frac{1}{2}<q<\frac{1}{2}$ Eq.~\eqref{small_q} should be multiplied by the corrective factor 
\begin{equation}
c_{\mathrm{cor}}(q)=1+\frac{\pi^{1-q}\, \epsilon^{2-4q}\, \Gamma(q-1/2)}{2^{1-2q}\, \Gamma(q)\, \Gamma(1/2-q)} N^{-(\gamma-1)(1-2q)}\ .
\label{correction}
\end{equation}
 The moment with $q=\frac{1}{2}$ is unusual  and contains an additional logarithm of $N$
 \begin{equation}
I _{\tfrac{1}{2}}=N^{1-\gamma/2}\, C_{\frac{1}{2}},\; C_{\frac{1}{2}}=\frac{\epsilon}{\pi}\Big[2(\gamma-1)\ln N-\ln\big (\frac{\pi \epsilon^4}{16}\big )-\gamma  \Big ]
\label{q_05}
 \end{equation}
 where $\gamma\approx 0.5772$ is the Euler constant. 
   
\begin{figure}
  \begin{center}
 \includegraphics[width=.99\linewidth]{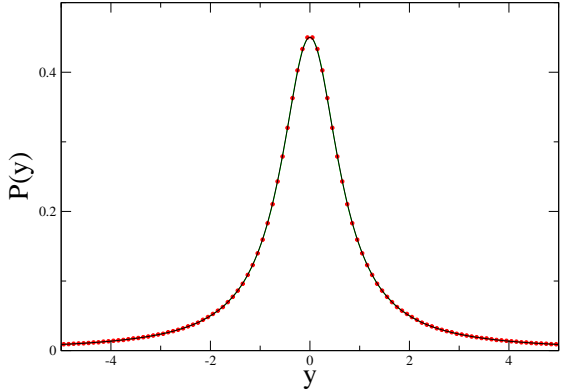}
  \end{center} 
 \caption{Red circles: distribution of $y=N^{\gamma/2}\ \Psi_j(E)$ for the RP model with parameters $\gamma=1.5$, $\epsilon=\frac{1}{\sqrt{2}}$ in the bulk computed numerically for $N=4096$. Data for $N=1024$ and $N=2048$ are indistinguishable  from the ones with $N=4046$. The solid black line is the theoretical prediction for this quantity \eqref{bulk}.} 
 \label{fig_bulk}
 \end{figure}

\begin{figure}
  \begin{center}
 \includegraphics[width=.99\linewidth]{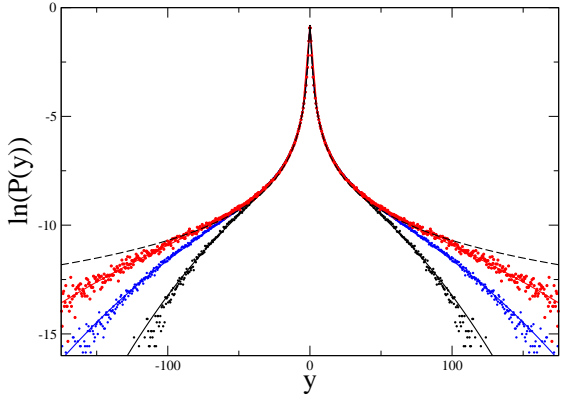} 
  \end{center} 
 \caption{The same as in Fig.~\ref{fig_bulk} but in the logarithmic scale. Black points: $N=1024$, blue points: $N=2048$, red points: $N=4096$. The solid lines of the same colour are theoretical predictions given by Eq.~\eqref{main} with $C=N^{\gamma/2}$. The dashed  black line  is the logarithm of the bulk Cauchy distribution \eqref{bulk}.} 
 \label{fig_tail}
 \end{figure}

  To illustrate the precision of the obtained formulas we calculate numerically the eigenfunction distribution for the RP model \eqref{matrix}  with $\gamma=1.5$ and $\epsilon=1/\sqrt{2}$  and matrix dimensions $N=1024, 2048$ and  $4096$. We  take into account eigenfunctions with eigenvalues in one eighth of the spectrum around the centre.  For the first two values of $N$ 10000 different realisations of random matrices were performed and for $N=4096$ the number of realisations was 1000.  The results are presented at Figs.~\ref{fig_bulk} and \ref{fig_tail}.  We checked that the distribution does not depend on the component $j$ and averaged over a few components. The agreement of the numerical results with the theoretical predictions is excellent.

\begin{figure}
 \begin{center}
\includegraphics[width=.99\linewidth]{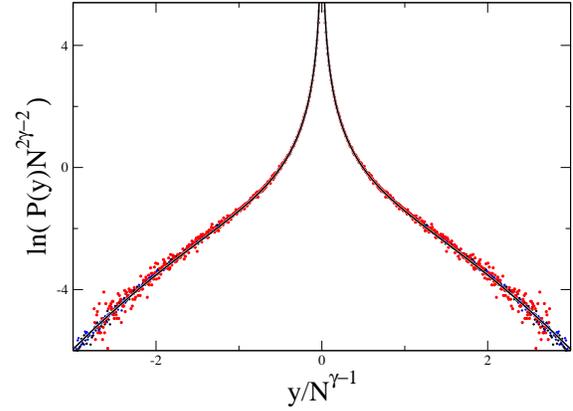}
 \end{center} 
 \caption{The same as in Fig.~\ref{fig_tail} but rescaled as indicated. The black line  is the logarithm of the tail  distribution \eqref{tail} without the factor $N^{\gamma-1}$.} 
 \label{fig_rescaling}
 \end{figure} 
 
 \begin{figure}
 \begin{center}
 \includegraphics[width=.99\linewidth]{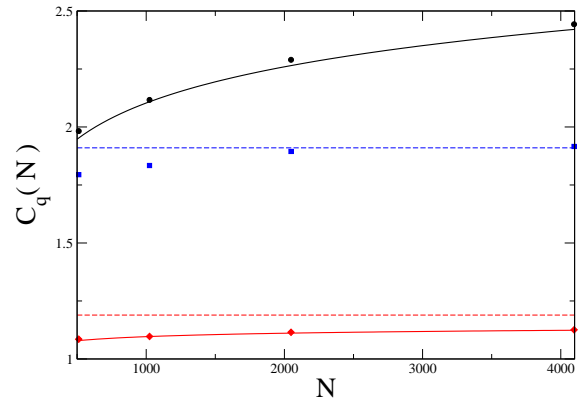}
 \end{center} 
 \caption{Eigenfunction moments with $q=\frac{1}{2}$ (black circles),  $q=2$ (blue squares), and $q=\frac{1}{8}$ (red diamonds) calculated numerically for the RP model with the same parameters as in Fig.~\ref{fig_bulk}  for $N=512,\ 1024,\ 2048,\ 4096$ divided by the corresponding powers of $N$.  Black solid line is Eq.~\eqref{q_05}  including the logarithmic term. Dashed blue and red lines  indicate the constant asymptotic values calculated from Eqs.~\eqref{large_q}, and \eqref{small_q}: $C_{2}=1.91$ and $C_{\tfrac{1}{8}}=1.19$. Solid red line shows the correction term \eqref{correction} which for the chosen parameters is $c_{\mathrm{cor}}\approx 1.19(1-.44/N^{1/4})$. } 
 \label{fig_moments}
 \end{figure} 
 
In Fig.~\ref{fig_rescaling} the data of Fig.~\ref{fig_tail} were rescaled to investigate the region of large eigenfunction values. The data for different $N$ are completely superimposed and agree very well with Eq.~\eqref{tail}. 
 
 In Fig.~\ref{fig_moments} numerically calculated moments are compared with Eqs.~\eqref{small_q}, \eqref{large_q}, and \eqref{q_05}.  In all considered cases numerical data agree very well with theoretical predictions. As higher moments are determined by the tail of the distribution (i.e. by rare events), their accurate numerical determination requires  a large number of realisations.  
 
 In the localised phase when $\gamma>2$ the same formulas remain valid. The main difference with the case $1<\gamma<2$ is that the large $x$ expansion \eqref{tail} does not exist due to the restriction $|\Psi_j|\leq 1$ as has been mentioned above. Consequently, the eigenfunction distribution is given by the Cauchy expression \eqref{bulk} which is sharply cut at the maximal possible value \eqref{xn} (which corresponds to strong localisation).  It means that eigenfunction moments are given by Eq.~\eqref{small_q} provided that $\gamma q<1$. All higher moments are determined by values $|\Psi_j|\sim 1$ which implies that higher fractal  dimensions are zero in agreement with \cite{kravtsov}.

 In conclusion, we have derived the statistical distribution for eigenfunctions of the Rosenzweig-Porter model in the regime $1<\gamma<2$. Our calculations are based on two well accepted physical assumptions. The first states that the mean square modulus of eigenfunctions  is given by the Breit-Wigner formula with the spreading width, $\Gamma$,  calculated by the Fermi golden rule. The second stipules that the eigenfunctions are distributed according to a local Porter-Thomas law with the variance given by the above formula. The final result is obtained by the averaging over diagonal matrix elements. This approach  is very simple, based on robust ideas, and leads to transparent explicit formulas which agree extremely well with numerical calculations.  Our results fully support the qualitative findings of \cite{kravtsov}  but have the advantage that all calculations are exact  and practically  all quantities can be obtained in closed form for large but  finite matrix dimensions.  
\begin{acknowledgments}
The authors are greatly indebted to J. Keating, J. Marklof, and Y. Tourigny for many useful discussions, to S. Warzel for pointing out Ref.~\cite{benigni}
and to V. Kravtsov for careful reading of the manuscript.
One of the authors (EB)  is grateful to  the Institute of Advance Studies at the University of Bristol for financial support in form of a Benjamin Meaker Visiting Professorship and to the School of Mathematics for hospitality during the visit where this paper was written.   
\end{acknowledgments}

\end{document}